\documentclass[12pt]{iopart}

\usepackage{graphicx}  
\begin{document}

\title{High--precision spectra for dynamical Dark Energy cosmologies
from constant--$w$ models}

\author{ Luciano Casarini}

\address{Department of Physics G.~Occhialini -- Milano--Bicocca
 University, Piazza della Scienza 3, 20126 Milano, Italy \&
I.N.F.N., Sezione di Milano}

\begin{abstract}
Spanning the whole functional space of cosmologies with any admissible
DE state equations $w(a)$ seems a need, in view of forthcoming
observations, namely those aiming to provide a tomography of cosmic
shear. In this paper I show that this duty can be eased and that a
suitable use of results for constant--$w$ cosmologies can be
sufficient.  More in detail, I ``assign'' here six cosmologies, aiming
to span the space of state equations $w(a) = w_o + w_a(1-a)$, for
$w_o$ and $w_a$ values consistent with WMAP5 and WMAP7 releases and
run N--body simulations to work out their non--linear fluctuation
spectra at various redshifts $z$. Such spectra are then compared with
those of suitable {\it auxiliary} models, characterized by constant
$w$. For each $z$ a different auxiliary model is needed. Spectral
discrepancies between the assigned and the auxiliary models, up to $k
\simeq 2$--$3\, h\, $Mpc$^{-1}$, are shown to keep within $1\, \%$.
Quite in general, discrepancies are smaller at greater $z$ and exhibit
a specific trend across the $w_o$ and $w_a$ plane. Besides of aiming
at simplifying the evaluation of spectra for a wide range of models,
this paper also outlines a specific danger for future studies of the
DE state equation, as models fairly distant on the $w_0 $--$w_a$ plane
can be easily confused.
\end{abstract}


\maketitle
\section{Introduction}
\label{sec:intro}

One of the main puzzles of cosmology is why a model as $\Lambda$CDM,
with so many conceptual problems, fits data so nicely.  It is then
important that the fine tuning paradox of $\Lambda$CDM is eased, with
no likelihood downgrade \cite{colombo,lavacca}, if Dark Energy (DE) is
a self--interacting scalar field $\phi$ (dDE cosmologies).

Although several researchers privilege potentials allowing tracking
solutions \cite{DDE1,DDE2}, data on $V(\phi)$ can be recovered just by
testing the evolution of the DE scale parameter, $w(a)$. Here $a =
1/(1+z)$ is the scale factor in the spatially flat metric $ ds^2 = c^2
dt^2 - a^2(t) d\ell^2~, $ with $d\ell$ being the comoving spatial
distance element.

\begin{figure}[htbp]
\begin{center}
\includegraphics[width=12cm]{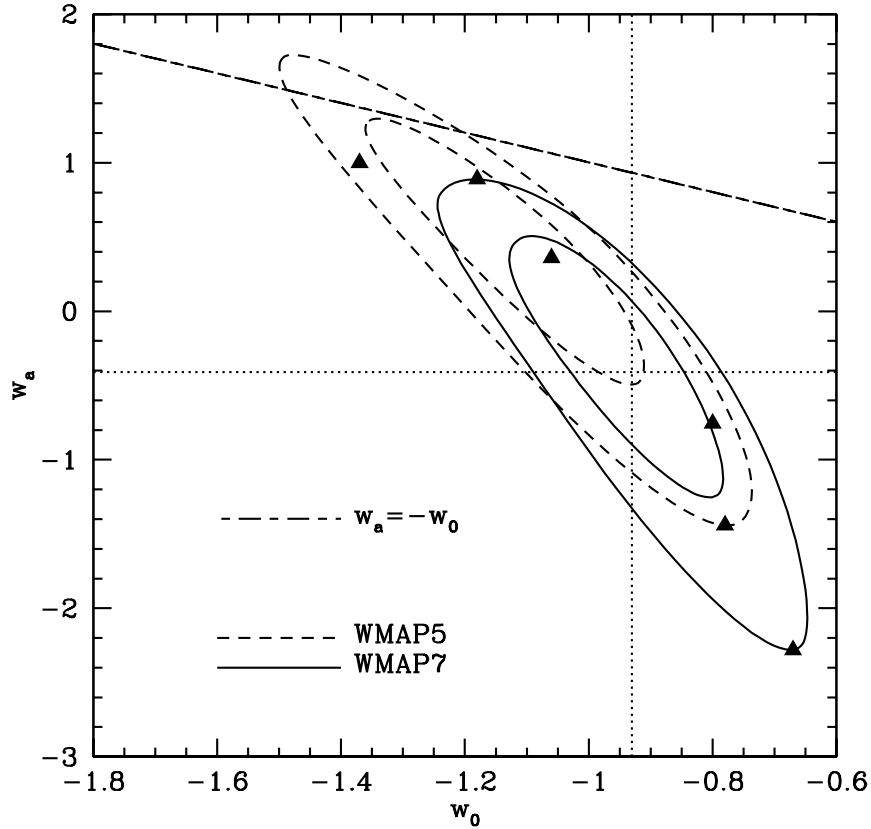}
\end{center}
\caption{Likelihood ellipses on the $w_o$--$w_a$ plane from WMAP5 and
WMAP7 data releases. The black triangles indicate the $\cal A$ models
considered in this work. State equations beyond the $w_o = -w_a$ line
should be modified at high--$z$. The dotted lines cross on the model
best fitting recent data.}
\label{ellipse2}
\end{figure}

Using available data, the WMAP team \cite{komatsu1, komatsu2} tried
to constrain the coefficients $w_0$ and $w_a$ in the expression
\begin{equation}
w(a) = w_0 + (1-a)\, w_a
\label{wa}
\end{equation}
for the DE state parameter. However, the setting of the likelihood
ellipse suggested, on the $w_o$--$w_a$ plane, has significantly
changed from WMAP5 to WMAP7 release. 
 The two ellipses are overlapped in Figure
\ref{ellipse2}, where we also indicate the boundary line $w_o = -w_a$,
beyond which the DE state equation should be rejected, unless
further modified by other parameters at high $z$ (otherwise, DE
could become too dense, possibly modifying BBN and even Meszaros'
effect).

A further selection among models in such ellipses shall be provided,
in a near future, by different observations ({\it e.g.}
\cite{mota1,mota2}) and, in particular, by tomographic shear surveys ({\it e.g.}
\cite{refregier}), able to reconstruct the matter fluctuations spectra $P(k,z)$ at various
$z$'s, with a precision approaching 1$\, \%$ \cite{huterer}. It is
therefore important to provide a tool to ease the determination of
model spectra with complex DE state equations; namely avoiding the
need to explore the whole functional space of $w(a)$.

Quite in general, such spectra are to be obtained through N--body
simulations; for cosmologies with a variable state parameter $w(a)$,
they have been performed since 2003 ({\it e.g.} \cite{klypinM},\cite{linderW}, \cite{maccio},\cite{solevi}) and compared with $w= const$ simulation
outputs. Observables considered in these papers, however, only
marginally included spectra.

An important step forward was then due to Francis et
al. \cite{francis}. They showed that suitable tuned constant--$w$
models, at $z=0$, closely approximate the spectra of cosmologies with
a state parameter given by eq.~(\ref{wa}). More precisely, the
spectrum $P(k,0)$ of an {\it assigned} ($\cal A$) model with state
parameter $w(a)$, can be approached by an {\it auxiliary}
constant--$w$ model ($\cal W$) such that: (i) in $\cal A$ and $\cal
W$, $\Omega_{b,m,tot}$, $h$ and $\sigma_8$ are equal, (ii) the
constant DE state parameter of $\cal W$ is tuned to yield equal
comoving distances from the Last Scattering Band (LSB) and $z=0$, for
$\cal W$ and $\cal A$. Then spectral discrepancies keep $< 1\, \%$, up
to $k \sim 2$--$3\, h\, $Mpc$^{-1}$ (here symbols have their usual
meaning).

Discrepancies between $\cal A$ and $\cal W$, at higher $z$, were also
tested in \cite{linder}, but they increase up to several~percents, so
that their precision may not be enough to exploit data.

The required precision was then obtained through a technique
introduced by Casarini et al. \cite{CaMaB} (Paper I hereafter) and
tested for two specific cosmologies (see also \cite{hydro}, for a
further extension based on hydrodynamical simulation). Here I plan to
test more cosmologies, so to sample the parameter space compatible
with WMAP5 and WMAP7 data, also exploring the precision trend in its
different regions. The setting of the six $\cal A$ models considered
is shown in Figure \ref{ellipse2}.
The other parameters are consistent with both WMAP5 and WMAP7 data: matter density $\Omega_m=0.274$, Hubble parameter [100 km/s/Mpc] $h=0.7$, fluctuation amplitude at $8h^{-1}$Mpc $\sigma_8=0.81$ and scalar spectral index $n_s=0.96$.

The plan of the paper is as follows:
in \S 2 is described the approach used in Paper I, \S 3 is devoted to describing our simulations and the techniques used to analyse them, in \S 4 are presented our results, and in \S 5 are discussed them. In Appendix A is reported  the algebraic technique used to reproduce ellipses in Figure \ref{ellipse2}.

\section{The spectral equivalence criterion}
\label{}
Let me then first recall the technique presented in Paper I. At
variance from \cite{linder}, given an assigned model $\cal A$, we
introduced a specific auxiliary model $\cal W$$(z)$, for each $z$;
$\cal A$ and $\cal W$$(z)$, first of all, are required to share the
values of $\omega_{b,c,m} = \Omega_{b,c,m} h^2$ and~$\sigma_8$, at
such$~z~.$

The former request is easily fulfilled; in fact, at any redshift the
critical density $\rho_{cr}$ is defined through the value of the
Hubble parameter $H$, being $ H^2 = (8\pi G/3) \rho_{cr} $. If we
multiply both sides of this relation by $\Omega_{m}$ (or
$\Omega_b,~\Omega_c$) we have
\begin{equation}
\Omega_m H^2 = (8\pi G/3) \rho_{m}~.
\end{equation}
The r.h.s. of this equation, and then $\omega_m \propto \Omega_m H^2$
(or $\omega_b,~\omega_c$), scale as $a^{-3}$, independently of the
model. Accordingly, once $\cal A$ and $\cal W$ share $\omega_{b,c,m}$
at $z=0$, it is so at any $z$: all $\cal W$$(z)$ models have equal
$\omega_{b,c,m}$.

On the contrary, the evolution of $\sigma_8$ depends on DE state
equation. Its value at $z=0$, as well as the value needed to normalize
initial conditions, can be worked out only once we know the {\it
constant} DE state parameters $w(z)$ of the $\cal W$$(z)$ models.

We come then to the most specific requirement, causing the dependence
on $z$ of the constant $w$'s: that $w$ is tuned so that $\cal W$$(z)$
and $\cal A$ have equal comoving distances between $z$ and the LSB.

The choice of $H_o$ (the Hubble parameter at $z=0$) is still
unconstrained. Taking it however equal to $H_o$ in $\cal A$ yields
boxes with equal side $L$ in both Mpc and $h^{-1}$Mpc units. Notice
that a simple--minded generalization of the criterion in
\cite{linder}, to high $z$, requires equal $\Omega_{b,c,m}(z)$ and,
thence, $H(z)$; this would create serious problems of sample variance
and model comparison.

\section{Simulations}

Simulations performed for this work are meant to test the spectra of
the $\cal M$ models against the corresponding auxiliary $\cal W$
models up to $z = 2$. We compare simulations starting from
realizations fixed by using an identical random seed. Initial
conditions have been created, at $z=24$, with the same procedure as in
Paper I. They were then run by using the {\sc pkdgrav}
code~\cite{stadel}, modified to deal with any variable $w(a)$ for
Paper I. All models are run in a box with side $L_{box}=256
h^{-1}$Mpc, using $N=256^3$ particles and a gravitational softening
$\epsilon = 25 h^{-1}$ kpc.

Besides of the six models $\cal A$, we have 4 auxiliary models $\cal
W$$(z)$ for $z = 0,~0.5,~1,~2~,$ which were run just down to the
redshift where they are tested. Altogether, therefore, we run 30 model
simulations.

Model spectra are then worked out through a Fast Fourier Transform
(FFT) of the matter density field. This last quantity is computed on a
regular grid $N_G\times N_G\times N_G$ (with $N_G = 2048$) from the
particle distribution via a Cloud in Cell algorithm.

Mass functions were also worked out for all models and found to be
consistent with Sheth \& Tormen predictions.

\begin{figure}[htbp]
\begin{center}
\includegraphics[width=15cm]{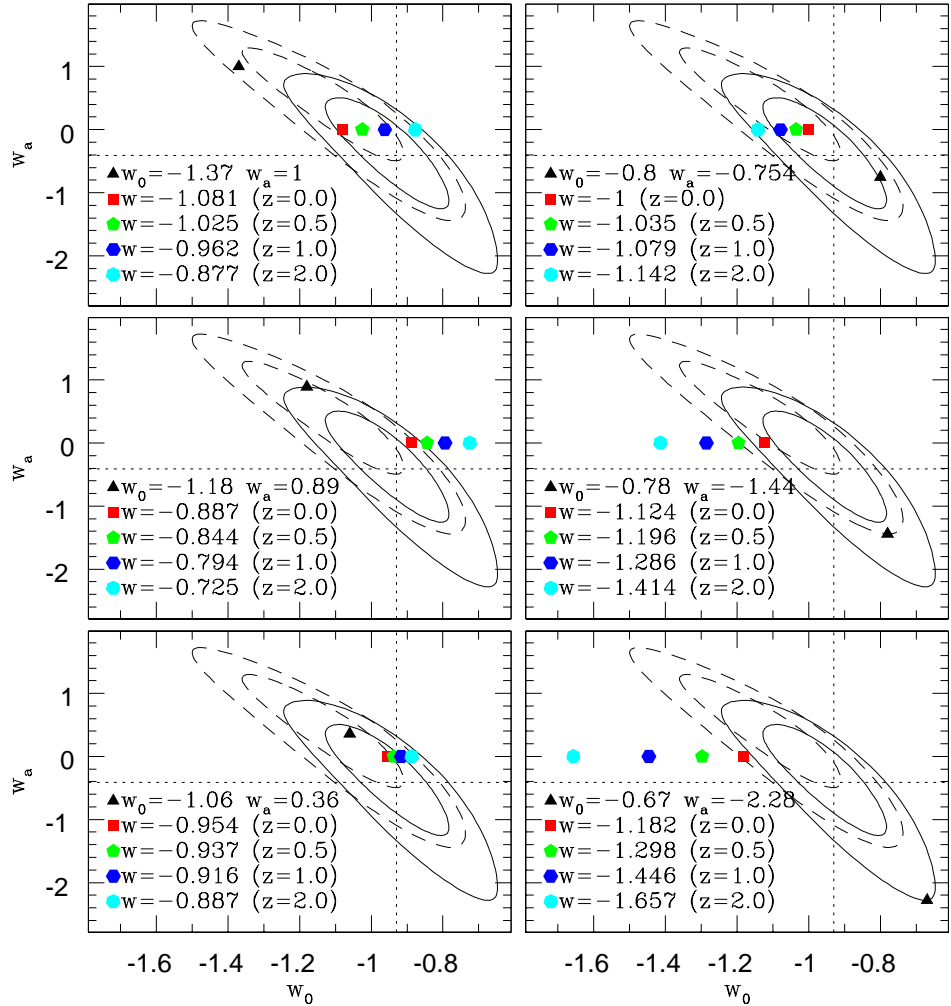}
\end{center}
\caption{Each box in this Figure refers to a single $\cal A$ model
(black triangle) and the related $\cal W$$(z)$ models (color
polygones). Notice that, in most cases, distances between colored
polygones are smaller than their distance from the black triangle.
This outlines the possibility of a serious bias in data analysis,
if tested by assuming constant $w$ cosmologies.}
\label{Fe}
\end{figure}
\begin{figure}[htbp]
\begin{center}
\includegraphics[width=15cm]{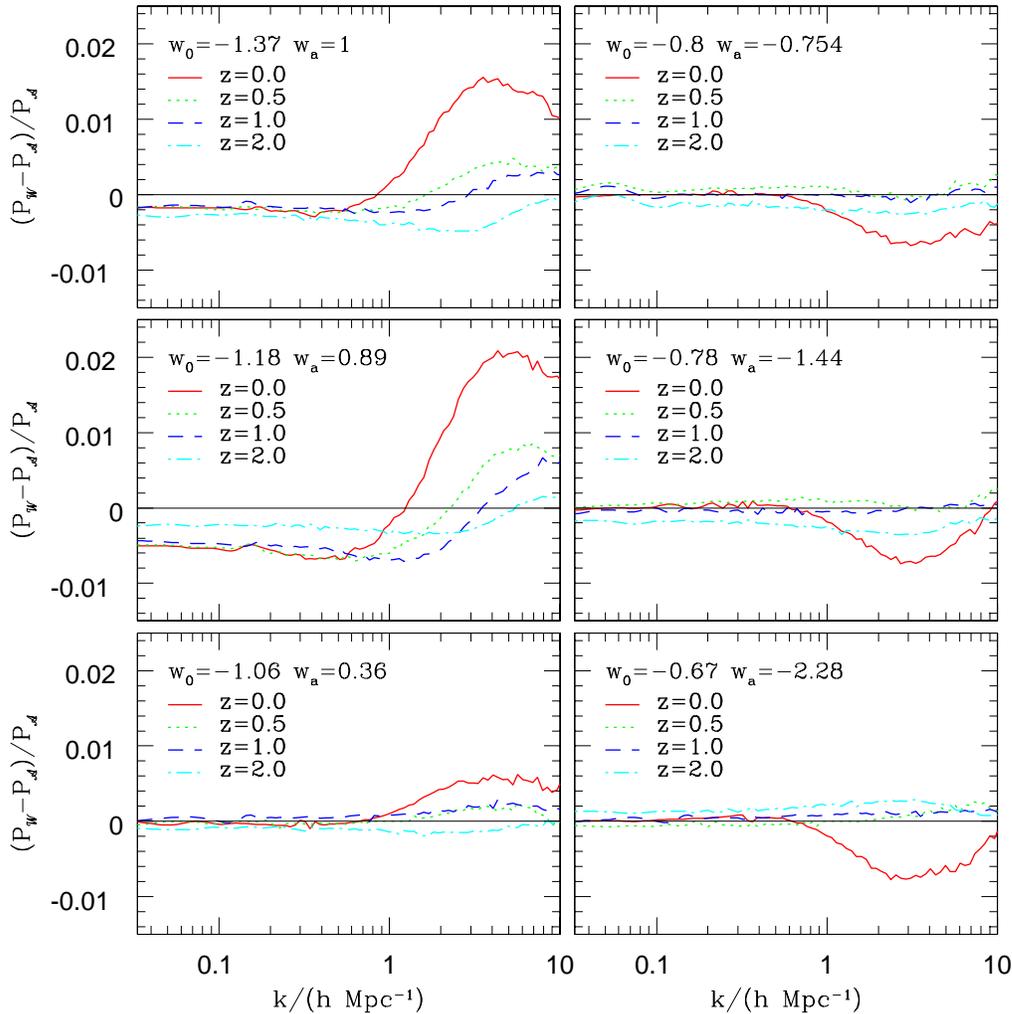}
\end{center}
\caption{Spectral discrepancies at various redshift. Each box refers to
a single $\cal A$ model (same ordering as in previous Figure).
Discrepancies are however greatest at $z=0$ and, at such redshift
only, attain $2 \, \%$ in models with $w_o \ll -1$. The $k$ value
where such large discrepancy is reached, however, is $\sim 4$--$5\,
h\, $Mpc$^{-1}$, well above the $k$--range where gas dynamics can be
ignored, if aiming at $\sim 1\, \%$ precision. Keeping within $k
\simeq 4$--$5\, h\, $Mpc$^{-1}$, the top discrepancy is always $< 1\,
\%~.$
  }
\label{Fa}
\end{figure}

\section{Results}
\label{}

In what follows, $\cal A$ models will be ordered according to
increasing values of the parameter $w_o~,$ from $\cal A$$1$ to $\cal
A$$6$. In Figure \ref{Fe} I report the setting of $\cal A$'s and
related $\cal W$--models on the $w_o$--$w_a$ plane, for
$z=0,~0.5,~1,~2~$.

The $\cal A$$1$ cosmology was consistent with WMAP5 and is apparently
outside the 2--$\sigma$ curve for WMAP7. It should be however reminded
that the significant shift of the ellipses is unlikely due to the fresh CMB inputs, while omitting to impose the distance prior
\cite{komatsu1,komatsu2} surely had an impact on it.

Figure \ref{Fe} indicates that the distance between $\cal W$ models,
lying on the $w_a=0$ line, by definition, is mostly smaller than their
distance from $\cal A$. Distances however scale with $w_o$ and are
smaller for the central $w_o$ values. I shall return on this point in
the next Section.

Figure \ref{Fa} then shows spectral discrepancies. Within $k = 3\, h\,
$Mpc$^{-1}$ the maximum discrepancy is reached for $\cal A$$1$ and
$\cal A$$2$ at $z=0$, attaining 1.4 and 1.6$\, \%$,
respectively. Owing to \cite{hydro}, however, this is a scale where
baryon dynamics affects spectra already more than 1--2$\, \%$. In
Paper I we actually took $k = 3\, h\, $Mpc$^{-1}$ as a limit; in this
scale range, however, the discrepancy is rapidly bursting and, should
we keep within $k = 2\, h\, $Mpc$^{-1}$, no spectral discrepancy
exceeds 1$\, \%$.

It is however clear that the other models exhibit a nicer behavior.
Even the $\cal A$$6$ model, whose distance from $\Lambda$CDM is
similar to $\cal A$$1$, exhibits discrepancies within $0.8\, \%$, up to
$k=10~.$ Other models are even nicer, exhibiting discrepancies in the
range of the permil.

Discrepancies decrease with increasing $z$ and, at $z = 0.5$ are
mostly in the permil range, for all models.

\section{Discussion}

This finding perfectly fits the reason why spectral similarities are
expected. When using the conformal time $\tau$, the background cosmic
metric reads $ds^2 = a^2 (\tau) (d\tau^2 - d\ell^2)$. Equal comoving
distance means then that equal conformal times have elapsed. Most
events, on a cosmological scale, are indeed scheduled in accordance
with the conformal time $\tau~$; the ordinary time $t$, instead, sets
a correct ``timing'' in virialized environments, where local
minkowskian reference frames no longer feel the scale factor
evolution.  Accordingly, it makes sense to set aside models with equal
``conformal age'', expecting increasing discrepancies when time
elapses and, however, on scales virialized since longer. Up to $k \sim
2$--$3\, h\, $Mpc$^{-1}$ we are inspecting scales $>\sim L=2\pi/k \sim
2\, h^{-1}$Mpc; still at $z=0$, there are quite a few virialized
systems on such scales, which however correspond to density peaks more
and more above average, as we go towards higher $z~$.

A further comment is deserved by the proximity of auxiliary models on
the $w_o$--$w_a$ plane. In quite a few cases, as for the models $\cal
A$$3$ and $\cal A$$4$, they almost overlap. This envisages a danger,
in future observational analysis; it is reasonable to expect that a
first data test is carried by assuming constant $w$. Let us suppose,
for instance, that the real cosmology is close to the model $\cal
A$4. It is not unlikely that all the models indicated by colored
polygones are then compatible with a single $w$ value, {\it e.g.}
-1.18$\pm 0.10~.$

As a matter of fact, starting from the setting of each triangle, we
could draw a bunch of curves, indicating the {\it loci} of equal
$\tau_o-\tau_{rec}$ (difference between conformal present and
recombination times), when $w_a$ varies. When these curves diverge
fast enough, there is a realistic possibility that they cross the
$w_a = 0$ line on reasonably distant sites. Otherwise, the risk of
spurious constant--$w$ detection is a serious danger. One should also
take into account that the very assumption of a polynomial $w(a)$ is a
simplifying ansatz, and that there can be cosmologies behaving even
more dangerously, {\it e.g.} some cosmology arising from a tracking
potential.

In this context, the very discrepancies detected above $k \sim
2$--$3\, h\, $Mpc$^{-1}$ are welcome. These scales still need to be
tested through hydro simulations, but a reliable pattern for data
analysis could actually start from the assumption of constant--$w$, so
individuating a bunch of curves, characterized by constant
$\tau_o-\tau_{rec}$, which will be the models among which one will
discriminate through higher--$k$ spectral discrepancies.

\begin{ack}

Luca Amendola and Loris Colombo are gratefully thanked for their
comments on this work. A particular thank is also due to Silvio
Bonometto, for his suggestions and advises during the preparation of
this article. Work partially supported by ASI (Italian Space Agency)
through the COFIS program.

\end{ack}

\appendix
\section{Reproduction of the likelihood ellipses with the {\it B\'ezier} curves}
The algebraic technique
used to reproduce the ellipses in Figure \ref{ellipse2} is named
after {\it B\'ezier} and is largely used in vector graphics to model
smooth curves which can be scaled indefinitely, without any bound, by
the limits of rasterized images. In the PostScript files in the WMAP
$\Lambda$--site, I found the coefficients for the cubic B\'ezier
curves:
\begin{equation} 
\label{bezier}
{\bf B}(u)=(1-u)^3 {\bf P}_0 + 3(1-u)^2 u {\bf P}_1+3 (1-u)u^2 {\bf
P}_2+u^3 {\bf P}_3~~ u \in [0,1]
\end{equation}
yielding 1-- and 2--$\sigma$ contours. Here the vector {\bf B},
running on the $w_0$--$w_a$ plane, describes a curve fixed by the
positions of the points $P_k$ (k=0,...,3), when $u$ varies from 0 to 1. Further details on this technique can be found in a previous paper \cite{casarini}, where the coordinates of B\'ezier points for WMAP5 ellipses are also reported. 
Here below I report the coordinates of the B\'ezier
points to draw WMAP7 ellipses.

\begin{table}[htbp]
\begin{center}
\begin{tabular}{|l||c|c||c|c||c|c||c|c||}
\hline
  & $x_0$ & $y_0$ & $x_1$ & $y_1$ & $x_2$ & $y_2$ & $x_3$ & $y_3$\\
\hline
i & & & & & & & & \cr
\hline
\hline
1 & -1.1203 & 0.4568 & -1.0857 & 0.6227 & -0.9828 & 0.3566 & -0.8997 & -0.0810\\
\hline
2 & -0.8997 & -0.0810 & -0.8166 & -0.5186 & -0.7648 & -1.0959 & -0.7890 & -1.2234\\
\hline
3 & -0.7890 & -1.2234 & -0.8132 & -1.3508 & -0.9069 & -1.0583 & -1.0034 & -0.5500\\
\hline
4 & -1.0034 & -0.5500 & -1.0999 & -0.0416 & -1.1483 & 0.3225 & -1.1203 & 0.4568\\
\hline
\end{tabular}
\end{center}
\vskip -1.truecm
\label{tsig1}
\end{table}
\begin{table}[htbp]
\begin{center}
\begin{tabular}{|l||c|c||c|c||c|c||c|c||}
\hline
 & $x_0$ & $y_0$ & $x_1$ & $y_1$ & $x_2$ & $y_2$ & $x_3$ & $y_3$ \cr
\hline
i &  &  &  &  &  &  &  & \\
\hline
1 & -1.2385 & 0.7974 & -1.1949 & 1.0412 & -1.0062 & 0.8187 & -0.8278 & -0.1565\\ 
\hline
2 & -0.8278 & -0.1565 & -0.6626 & -1.0603 & -0.6270 & -2.1171 & -0.6565 & -2.2500 \\
\hline
3 & -0.6565 & -2.2500 & -0.6941 & -2.4197 & -0.8457 & -1.8601 & -1.0139 & -0.8592\\
\hline
4 & -1.0139 & -0.8592 & -1.1822 & 0.1417 & -1.2679 & 0.6336 & -1.2385 & 0.7974 \\
\hline
\end{tabular}
\end{center}
\caption{Points defining the 4 cubic B\'ezier expressions yielding the
1--$\sigma$ and 2--$\sigma$ curves (upper and lower table, respectively).}
\label{tsig2}
\end{table}

\newcommand{\Nature}{{\it Nature\/} }
\newcommand{\ApJ}{{\it Astrophys. J.\/} }
\newcommand{\ApJS}{{\it Astrophys. J. Suppl.\/} }
\newcommand{\MNRAS}{{\it Mon. Not. R. Astron. Soc.\/} }
\newcommand{\PhRv}{{\it Phys. Rev.\/} }
\newcommand{\PhL}{{\it Phys. Lett.\/} }
\newcommand{\JCAP}{{\it J. Cosmol. Astropart. Phys.\/} }
\newcommand{\AeA}{{\it Astronom. Astrophys.\/} }
\newcommand{\etall}{{\it et al.\/} }
\newcommand{\arXiv}{{\it Preprint\/} }
\newcommand{\NewA}{{\it New Astron.\/} }

{}

\end{document}